\documentclass[preprint,prd,nofootinbib,tightenlines,amsmath]{revtex4}
\usepackage{graphicx}
\usepackage{bm}

\begin{document}
\baselineskip=15pt \parskip=5pt

\vspace*{3em}

\title{The Simplest Dark-Matter Model, CDMS\,II Results, \\ and Higgs Detection at LHC}

\author{Xiao-Gang He$^{1}$}
\email{hexg@phys.ntu.edu.tw}
\author{Tong Li$^{2}$}
\email{allongde@mail.nankai.edu.cn}
\author{Xue-Qian Li$^{3}$}
\email{lixq@nankai.edu.cn}
\author{Jusak Tandean$^{1}$}
\email{jtandean@yahoo.com}
\author{Ho-Chin Tsai$^{1}$}
\email{hctsai@phys.ntu.edu.tw}

\affiliation{$^1$Department of Physics, Center for Theoretical Sciences, and LeCosPA Center, \\
National Taiwan University, Taipei 106, Taiwan \\
$^2$Center for High Energy Physics, Peking University, Beijing 100871, China \\
$^3$Department of Physics, Nankai University, Tianjin 300071, China}

\date{\today $\vphantom{\bigg|_{\bigg|}^|}$}

\begin{abstract}
The direct-search experiment for dark matter performed by the CDMS\,II Collaboration has
observed two candidate events.  Although these events cannot be interpreted as significant
evidence for the presence of weakly interacting massive particle (WIMP) dark matter (DM),
the total CDMS\,II data have led to an~improved upper-limit on the WIMP-nucleon
spin-independent cross-section.
We study some implications of these results for the simplest WIMP DM model, the SM+D, which
extends the standard model (SM) by the addition of a real SM-singlet scalar field dubbed
darkon to play the role of the DM.
We find that, although the CDMS\,II data rule out a sizable portion of parameter space of
the model, a large part of the parameter space is still allowed.
We obtain strong correlations among the darkon mass, darkon-nucleon cross-section,
mass of the Higgs boson, and branching ratio of its invisible decay.
We point out that measurements of the Higgs invisible branching-ratio at the LHC can lift
some possible ambiguities in determining the darkon mass from direct DM searches.
\end{abstract}

\maketitle

\section{Introduction}

It is now well established that 23\% of the energy density of the Universe is provided
by dark matter~\cite{Amsler:2008zz}.
Although the evidence for dark matter (DM) has existed for many decades,
the identity of its basic constituents has so far remained elusive.
One of the popular candidates for DM is the weakly interacting massive particle~(WIMP).
Indirect DM searches have turned up interesting results which may be interpreted
as evidence for WIMPs~\cite{deBoer:2008iu}, but it is very difficult to establish firmly
the connection to DM due to the indirect nature of the observed events.
Direct detection on the Earth is therefore crucial to determine the properties of DM.

A variety of experiments have been and are being carried out to detect DM directly by looking
for the recoil energy of nuclei caused by the elastic scattering of a WIMP off a nucleon.
Stringent bounds on the WIMP-nucleon elastic cross-section have been obtained from
the null results of such searches~\cite{Angle:2007uj,Ahmed:2008eu}.
The DAMA Collaboration has reported the observation of DM annual modulation signature~\cite{dama},
but no other experimental confirmation is yet available.
Very recently the CDMS Collaboration has completed their analysis of the final data runs of
the CDMS~II experiment and reported two candidate events~\cite{cdms-new}.
Although these events cannot be interpreted as significant evidence for WIMPs interacting with
nucleons, the new data combined with previous CDMS\,II results have led to the most stringent
upper-limit to date on the WIMP-nucleon spin-independent cross-section for WIMP masses larger
than 40\,GeV or so.  For instance, the WIMP-nucleon cross-section for a WIMP of mass 70\,GeV is
constrained to be smaller than $3.8\times10^{-44}$\,cm$^2$ at 90\% confidence
level~\cite{cdms-new}.
This result further restricts the parameter space of a given WIMP model~\cite{cdms-theory}.

To explain the existence of WIMP DM, the SM  must be extended.  The simplest model which has
a WIMP candidate is the SM+D, which extends the SM by the addition of a real SM-singlet scalar
field $D$, called darkon, to play the role of the DM.
This model was first considered by Silveira and Zee~\cite{Silveira:1985rk}.
A closely related model, with one or more SM-singlet complex scalars, was proposed by McDonald
several years afterwards~\cite{mcdonald}.
The darkon model was further explored later by other
groups~\cite{Burgess:2000yq,darkon,Bird:2004ts,He:2007tt,Barger:2007im,Davoudiasl:2004be,hexg1}.
In this work, we explore some implications of the new CDMS\,II results for the SM+D and also
study how measurements of the Higgs boson at the LHC can help reveal the darkon properties.
We show that this darkon model can provide a consistent interpretation of the CDMS\,II results,
with much of its parameter space not excluded by the data.
One important feature of the model is that it has a small number of parameters.
As we elaborate later, this gives rise to strong correlations between the darkon mass,
the darkon-nucleon cross-section, the mass of the Higgs boson, and the branching ratio of its
invisible decay.
The LHC, soon to be operating in full capacity, can thus offer complementary information
about the properties of the darkon.

\section{Brief description of SM+D and its relic density}

Before discussing our main results, we summarize some of the salient features of the SM+D.
Being a WIMP DM candidate, the darkon $D$ has to be stable, which can be realized by assuming
$D$ to be a~SM singlet and introducing a discrete $Z_2$ symmetry into the model.
Under the $Z_2$ transformation, \,$D\to-D$\, and all SM fields remain unchanged.
Requiring that the darkon interactions be renormalizable implies that $D$ can interact with
the SM fields only through its coupling to the Higgs-doublet field $H$.
It follows that the general form of the darkon Lagrangian, besides the kinetic part
\,$\frac{1}{2}\partial^\mu D\,\partial_\mu^{}D$\, and the SM terms, can be written
as~\cite{Silveira:1985rk,mcdonald,Burgess:2000yq}
\begin{eqnarray}  \label{DH}
{\cal L}_D^{} \,\,=\,\,
-\frac{\lambda_D^{}}{4}\,D^4-\frac{m_0^2}{2}\,D^2 - \lambda\, D^2\,H^\dagger H ~,
\end{eqnarray}
where  $\lambda_D^{}$,  $m_0^{}$, and $\lambda$  are free parameters.
The parameters in the potential should be chosen such that $D$ does not develop a vacuum
expectation value (vev) and the $Z_2$ symmetry is not broken, which will ensure that the darkon
does not mix with the Higgs field, avoiding possible fast decays into other SM particles.

The Lagrangian in Eq.~(\ref{DH}) can be rewritten to describe the interaction of the physical
Higgs boson $h$ with the darkon as
\begin{eqnarray} \label{ld}
{\cal L}_D^{} \,\,=\,\, -\frac{\lambda_D^{}}{4}\,D^4-\frac{\bigl(m_0^2+\lambda v^2\bigr)}{2}\,D^2
- \frac{\lambda}{2}\, D^2\, h^2 - \lambda v\, D^2\, h \,\,,
\end{eqnarray}
where  \,$v=246$\,GeV\,  is the vev of $H$,  the second term contains the darkon mass
\,$m_D^{}=\bigl(m^2_0+\lambda v^2\bigr){}^{1/2}$,\, and the last term,  \,$-\lambda v D^2 h$,\,
plays an important role in determining the relic density of the darkon.
It is clear that this model has a small number of unknown parameters: the Higgs and darkon masses
$m_h^{}$ and $m_D^{}$, respectively, the Higgs-darkon coupling $\lambda$, and the darkon
self-interaction coupling $\lambda_D^{}$.  In our analysis, $\lambda_D^{}$ will not be involved.

At leading order, for \,$m_D^{}<m_h^{}$\, the relic density results from the annihilation of
a darkon pair into SM particles through Higgs exchange~\cite{Silveira:1985rk,mcdonald,Burgess:2000yq},
namely \,$DD\to h^*\to X$,\, where $X$ indicates SM particles.
Since the darkon is cold DM, its speed is nonrelativistic, and so a darkon pair has an invariant
mass  \,$\sqrt s\simeq2m_D^{}$.\,  With the SM+D Lagrangian determined, the $h$-mediated
annihilation cross-section of a darkon pair into SM particles is then given
by~\cite{Burgess:2000yq}
\begin{eqnarray} \label{csan}
\sigma_{\rm ann}^{}\, v_{\rm rel}^{} \,\,=\,\,
\frac{8\lambda^2 v^2}{\bigl(4m_D^2-m_h^2\bigr)^2+\Gamma^2_h\,m^2_h}\,
\frac{\sum_i\Gamma\bigl(\tilde h\to X_i\bigr)}{2m_D} \,\,,
\end{eqnarray}
where  \,$v_{\rm rel}^{}=2\bigl|\bm{p}_D^{\rm cm}\bigr|/m_D^{}$\, is the relative speed of
the $DD$ pair in their center-of-mass (cm) frame, $\tilde h$  is a virtual Higgs boson having
the same couplings to other states as the physical $h$ of mass $m_h^{}$, but with an invariant
mass  \,$\sqrt s=2m_D^{}$, and \,$\tilde h\to X_i$\, is any possible decay mode of $\tilde h$.
To obtain \,$\Sigma_i\Gamma\bigl(\tilde h\to X_i\bigr)$,\, one computes
the $h$ width and then sets $m_h^{}$ equal to  $2m_D^{}$.
For \,$m_D^{}\ge m_h^{}$,\, darkon annihilation into a pair of Higgs bosons, \,$DD\to hh$,\,
also contributes to $\sigma_{\rm ann}^{}$, through $s$-, $t$-, and $u$-channel diagrams.
This would become one of the dominant contributions to $\sigma_{\rm ann}^{}$, along with
\,$DD\to h^*\to W^+W^-,ZZ$,\, if \,$m_D^{}\gg m_{W,Z,h}^{}$\,~\cite{mcdonald,Burgess:2000yq}.

For a given interaction of the WIMP with SM particles, its annihilation rate into the latter and
its relic density $\Omega_D^{}$ can be calculated and are related to each other by the thermal
dynamics of the Universe within the standard big-bang cosmology~\cite{Kolb:1990vq}.
To a good approximation,
\begin{eqnarray} \label{oh}
\Omega_D^{} h^2 \,\,\simeq\,\,
\frac{1.07\times 10^9\, x_f^{}}{
\sqrt{g_*^{}}\, m_{\rm Pl}\,\langle\sigma_{\rm ann}^{}v_{\rm rel}^{}\rangle\rm\,GeV} \,\,,
\hspace{2em}
x_f^{} \,\,\simeq\,\,
\ln\frac{0.038\,m_{\rm Pl}\,m_D^{}\,\langle\sigma_{\rm ann}^{}v_{\rm rel}^{}\rangle}{
\sqrt{g_*^{}\, x_f^{}}} \,\,,
\end{eqnarray}
where  $h$ is the Hubble constant in units of 100\,km/(s$\cdot$Mpc),\footnote{It is obvious that
this constant is not to be confused with the physical Higgs field, also denoted by~$h$.}
\,$m_{\rm Pl}^{}=1.22\times10^{19}$\,GeV\, is the Planck mass,
\,$x_f^{}=m_D^{}/T_f^{}$\, with $T_f^{}$ being the freezing temperature, $g_*^{}$ is the number
of relativistic degrees of freedom with masses less than $T_f^{}$, and
\,$\langle\sigma_{\rm ann}^{}v_{\rm rel}^{}\rangle$\, is the thermally averaged product of
the annihilation cross-section of a pair of WIMPs into SM particles and the relative speed of
the WIMP pair in their cm frame.

The current Particle Data Group value for the DM density is
\,$\Omega_D^{} h^2=0.113\pm0.003$\,~\cite{Amsler:2008zz}.
The very recent seven-year data from WMAP, combined with other data, have led to
the updated value~\,$\Omega_D^{} h^2=0.1123\pm 0.0035$\,~\cite{wmap7}.
Using this new number and Eq.~(\ref{oh}), one can restrict the ranges of $x_f^{}$ and
\,$\langle\sigma_{\rm ann}^{}v_{\rm rel}^{}\rangle$\,  as functions of WIMP mass $m_D^{}$
without knowing the explicit form of the SM-WIMP interaction~\cite{hexg1}.

A large range of the darkon mass values, from as low as hundreds of MeV to as high as
several TeV, has been considered in the
literature~\cite{Silveira:1985rk,mcdonald,Burgess:2000yq,darkon,Bird:2004ts,He:2007tt,
Barger:2007im,Davoudiasl:2004be,hexg1}.
As far as direct detection of the darkon is concerned, searches that are ongoing or to be carried
out in the near future are not expected to be sensitive to $m_D^{}$ values less than a~few
GeV~\cite{Angle:2007uj,Ahmed:2008eu,dama,cdms-new,Angloher:2002in,Lin:2007ka,Irastorza:2009qh}.
Moreover, for a~relatively light Higgs boson, with
\,$100{\rm\,GeV}\mbox{\footnotesize\,$\lesssim$\,}m_h^{}
\mbox{\footnotesize\,$\lesssim$\,}200$\,GeV,\,
earlier studies suggest that near-future searches may also have limited sensitivity to darkon
masses greater than 100~GeV~\cite{Davoudiasl:2004be,hexg1}.
For these reasons, in our numerical work in the next section we concentrate on the range
\,$5{\rm\,GeV}\le m_D^{}\le100{\rm\,GeV}$.

\section{Results and discussion}

\begin{figure}[ht]
\includegraphics[width=4in]{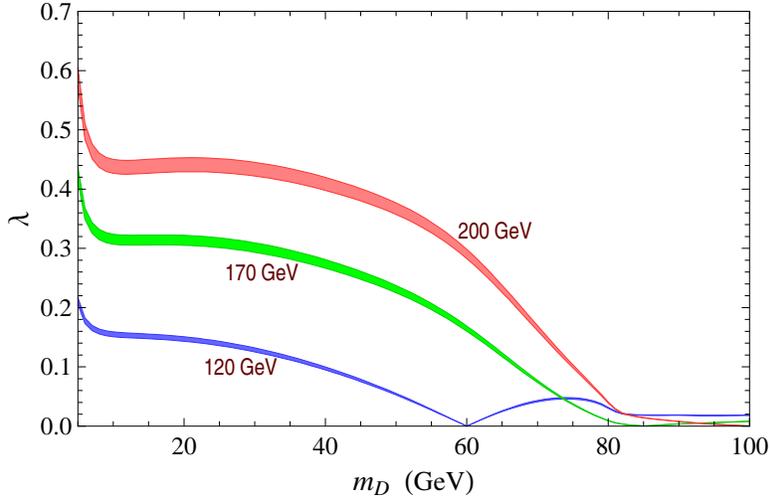}
\caption{Darkon-Higgs coupling $\lambda$ as a function of the darkon mass $m_D^{}$ for Higgs mass
values \,$m_h^{}=120,170,200$\,GeV.\,
The band widths in all figures result from the relic-density range which we have taken,
\,$0.1065\le\Omega_D^{}h^2\le0.1181$.\label{relic-md}}
\end{figure}

In the SM+D, one can draw a correlation among its parameters $\lambda$, $m_D^{}$, and
$m_h^{}$ from the range of \,$\langle\sigma_{\rm ann}^{}v_{\rm rel}^{}\rangle$\, values
allowed by the $\Omega_D^{}h^2$ constraint.
In this study, we adopt the \,$\langle\sigma_{\rm ann}^{}v_{\rm rel}^{}\rangle$\, range as
a~function of $m_D^{}$ obtained using the 90\%-C.L. range \,$0.1065\le\Omega_D^{}h^2\le0.1181$\,
derived from the new WMAP7 result quoted above~\cite{wmap7}.
To show the above-mentioned correlation, we plot in Fig.~\ref{relic-md} the allowed ranges
of $\lambda$ corresponding to \,$5{\rm\,GeV}\le m_D^{}\le100{\rm\,GeV}$\,  for some specific
values of the Higgs mass, which we choose to be \,$m_h^{}=120$, 170, and 200~GeV\,
for illustration.

One can make a few observations based on this figure.
First, it is clear that, although only a restricted range of the DM relic density is
allowed, it can be easily reproduced in this model.
Second, each of the dips of the curves corresponds to the minimum of the denominator in
Eq.~(\ref{csan}) at  \,$m_h^{}=2m_D^{}$.\,
Thus around this resonant point the interaction rate can be large even when $\lambda$ is small.
Third, $\lambda$ is not very small for the lower values of $m_D^{}$, and this will result in
sizable branching ratios of the Higgs invisible decay mode, as we will discuss further later.
Lastly, if in the near future it is the Higgs mass that is measured first (at the LHC)
rather than the darkon mass, one will have just one band to evaluate in order to probe
the darkon properties.

We remark here that, although the $\lambda$ values in Fig.~\ref{relic-md} tend to
become small as $m_D^{}$ approaches 100\,GeV, they can get large again,
approximately linearly with $m_D^{}$, if $m_D^{}$ is sufficiently large.
This follows from the facts that  \,$\langle\sigma_{\rm ann}^{}v_{\rm rel}^{}\rangle$\,
is roughly constant for the $m_D^{}$ range of interest and that
\,$\sigma_{\rm ann}^{}v_{\rm rel}^{}\simeq\lambda^2/\bigl(4\pi m_D^2\bigr)$\,
for \,$m_D^{}\gg m_{W,Z,h}^{}$\,~\cite{mcdonald,Burgess:2000yq,hexg1}

\begin{figure}[b] \vspace*{2ex}
\includegraphics[width=4in]{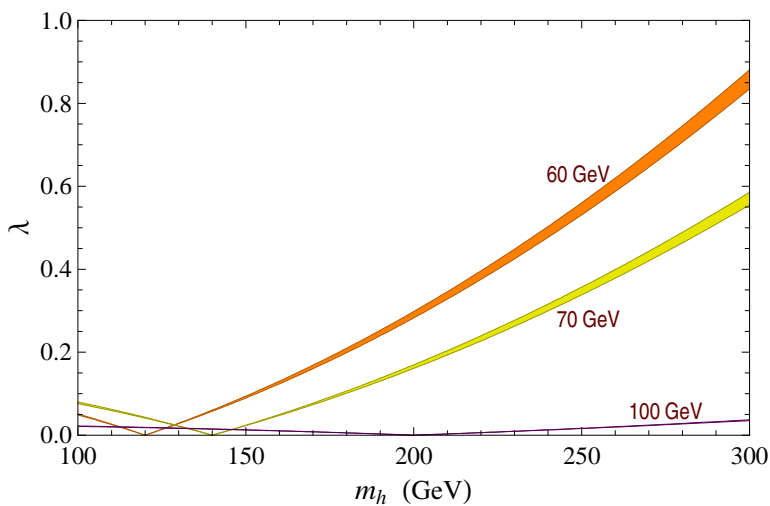}
\caption{Darkon-Higgs coupling $\lambda$ as a function of the Higgs mass $m_h^{}$ for darkon
mass values \,$m_D^{}=60,70,100$\,GeV.\label{relic-mh}}
\end{figure}

A complementary insight can be gained about the correlation between $\lambda$, $m_D^{}$, and
$m_h^{}$ from Fig.~\ref{relic-mh}, which displays the allowed ranges of $\lambda$ corresponding
to \,$100{\rm\,GeV}\le m_h^{}\le300{\rm\,GeV}$\,  for specific values of the darkon mass,
which we take to be \,$m_D^{}=60$, 70, and 100~GeV\,  for illustration.
Thus, if it is the darkon mass instead that is fixed first from a direct-search
experiment in the near future, one also needs to focus on only one band to study
the allowed ranges of the Higgs mass and the coupling $\lambda$.
We note again the resonant dips at \,$m_h^{}=2m_D^{}$.\,

Now, the direct detection of a WIMP on the Earth is through the recoil of nuclei when
the WIMP hits a~nucleon target.
Consequently, to make sure that the SM+D is consistent with the CDMS\,II data, we need to
check if the new bound on the darkon-nucleon  cross-section is satisfied.
In the SM+D, this interaction occurs via the exchange of a Higgs boson between the darkon
and the nucleon $N$ in the $t$-channel process \,$DN\to DN$,\,  which is in contrast to
the $s$-channel process of darkon annihilation in the \,$m_D^{}<m_h^{}$\, case.

To evaluate \,$DN\to DN$ requires knowing not only the darkon-Higgs coupling $\lambda$,
but also the Higgs-nucleon coupling $g_{NNh}^{}$, which parametrizes  the Higgs-nucleon
interaction described by  \,${\cal L}_{NNh}^{}=-g_{NNh}^{}\,\bar NN\,h$.\,
From this Lagrangian and ${\cal L}_D^{}$ in Eq.~(\ref{ld}), one can derive for
\,$|t|\ll m^2_h$\,  the darkon-nucleon elastic
cross-section~\cite{Silveira:1985rk,mcdonald,Burgess:2000yq,Barger:2007im,hexg1}
\begin{eqnarray} \label{csel}
\sigma_{\rm el}^{} \,\,\simeq\,\,
\frac{\lambda^2\,g_{NNh}^2\,v^2\,m_N^2}{\pi\,\bigl(m_D^{}+m_N^{}\bigr)^2\, m_h^4} \,\,.
\end{eqnarray}
In this approximation,  \,$\bigl(p_D^{}+p_N^{}\bigr){}^2\simeq\bigl(m_D^{}+m_N^{}\bigr){}^2$\,
has been used.

To compare with data, one then needs the value of $g_{NNh}^{}$, which is related to
the underlying Higgs-quark interaction described by
\,${\cal L}_{qqh}^{}=-\mbox{\large$\Sigma$}_q^{}m_q^{}\,\bar q q\,h/v$,\,
where the sum runs over the six quark flavors, \,$q=u,d,s,c,b,t$.\,
Since the energy transferred in the darkon-nucleon scattering is very small, of order a few
tens of keV, one can employ a chiral-Lagrangian approach to estimate~$g_{NNh}^{}$.
This has been done previously in the literature~\cite{Shifman:1978zn,Cheng:1988cz}.
More recently, we have also adopted this approach to estimate this coupling and
obtained~\cite{hexg1}
\begin{eqnarray}
g_{NNh}^{} \,\,\simeq\,\, 1.71\times10^{-3}  \,\,,
\end{eqnarray}
which is comparable to the values found in the literature~\cite{Burgess:2000yq,Cheng:1988cz}.
We will use this number in our numerical calculation.

With $\lambda$ being subject to the relic-density constraint and $g_{NNh}^{}$ known, we can
compute the darkon-nucleon elastic cross-section $\sigma_{\rm el}^{}$ as a function of darkon
mass for a fixed Higgs mass or as a~function of the Higgs mass for a fixed darkon mass.
A priori, the predicted $\sigma_{\rm el}^{}$ is not guaranteed to be compatible with the limits
from DM direct-search experiments.
Therefore, we have to check whether the SM+D prediction satisfies the new limit from CDMS\,II.

\begin{figure}[t]
\includegraphics[width=4in]{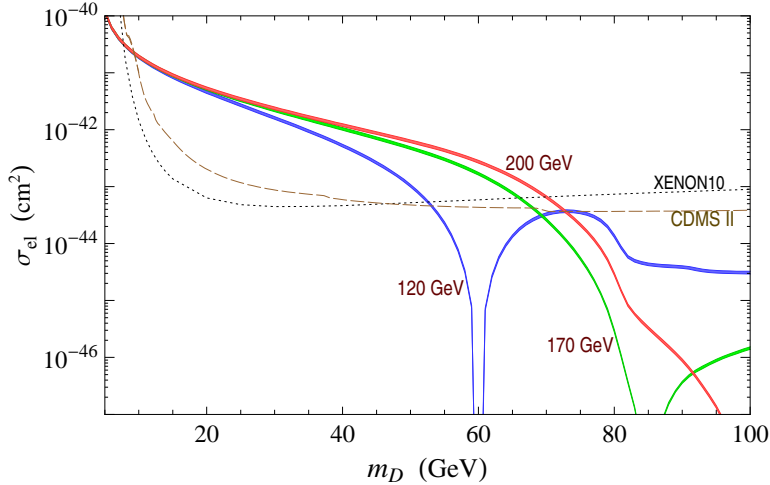} \vspace*{-1ex}
\caption{Darkon-nucleon elastic cross-section $\sigma_{\rm el}^{}$ as a function of the darkon
mass $m_D^{}$ for Higgs mass values \,$m_h^{}=120,170,200$\,GeV,\, compared to 90\%-C.L. upper
limits from CDMS\,II (dashed curve) and XENON10 (dotted curve).\label{cross-sec-md}}
\end{figure}

\begin{figure}[t] \vspace*{2ex}
\includegraphics[width=4in]{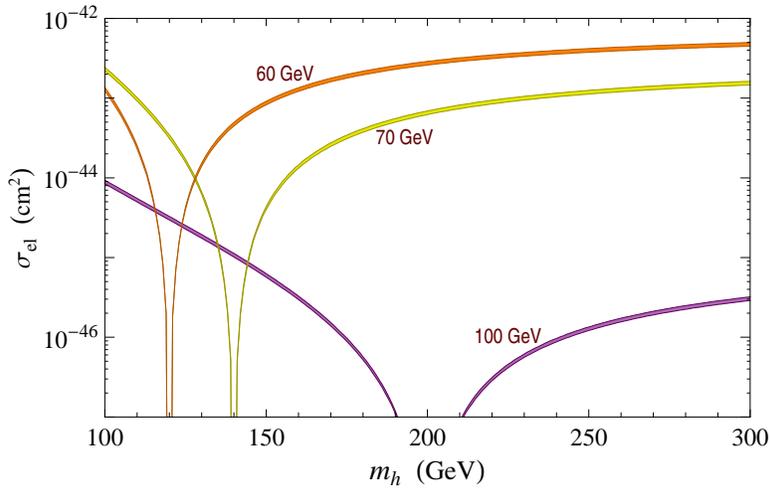} \vspace*{-1.5ex}
\caption{Darkon-nucleon elastic cross-section $\sigma_{\rm el}^{}$ as a function of
the Higgs mass $m_h^{}$ for darkon mass values \,$m_D^{}=60,70,100$\,GeV.\label{cross-sec-mh}}
\end{figure}

We show our results for $\sigma_{\rm el}^{}$ in Figs.~\ref{cross-sec-md} and~\ref{cross-sec-mh},
where the choices of Higgs and darkon masses are the same as those in Figs.~\ref{relic-md}
and~\ref{relic-mh}, respectively.
In Fig.~\ref{cross-sec-md}, we also plot the 90\%-C.L. upper-limit curve given in the new
CDMS\,II report~\cite{cdms-new}, as well as the corresponding limit set by the XENON10
experiment~\cite{Angle:2007uj}.
For \,$m_D^{}<10$\,GeV,\, there are additional constraints on the cross-section from
the CRESST-I~\cite{Angloher:2002in} and TEXONO~\cite{Lin:2007ka} experiments, but their limits
are of order $10^{-39}{\rm\,cm}^2$ or higher, exceeding the predictions.
In Fig.~\ref{cross-sec-mh}, to avoid cluttering the graph, we have not displayed
the experimental limits, as they depend on $m_D^{}$ and could be easily estimated
from~Fig.~\ref{cross-sec-md}.
We can see from Fig.~\ref{cross-sec-md} that there are significant regions in the SM+D parameter
space that are consistent with the CDMS\,II results, although a sizable portion of it is
not allowed by the current data.
More specifically, for \,$m_D^{}\mbox{\footnotesize\,$\lesssim$\,}7$\,GeV\, the model is viable,
to about \,$m_D^{}\sim2$\,GeV\, below which it is stringently constrained by the measured bounds
on $B$-meson decays into a $K^{(*)}$-meson plus missing energy~\cite{Bird:2004ts}, whereas for
higher $m_D^{}$ values the ranges that are allowed or ruled out depend on the Higgs mass.
For the \,$m_h^{}=120$, 170, and 200~GeV\, examples in the figure, $m_D^{}$ values from 8\,GeV
to about \,53, 68, and 73~GeV,\, respectively, are experimentally excluded.

Before proceeding, it is worth remarking that $\sigma_{\rm el}^{}$ for fixed $m_h^{}$
would approach a~constant value if \,$m_D^{}\gg m_{W,Z,h}^{}$.\,
This is because in this large-$m_D^{}$ limit the ratio \,$\lambda^2/m_D^2$\, is approximately
constant, as mentioned above, and $\sigma_{\rm el}^{}$ in Eq.~(\ref{csel}) is proportional
to the same ratio,~$\lambda^2/m_D^2$.
From Eq.~(\ref{csel}), one can also see that the asymptotic value of $\sigma_{\rm el}^{}$
decreases as $m_h^{}$ increases.

We now discuss a number of features worth pointing out.
If \,$m_h^{}=2m_D^{}$,\,  the relic density is determined at the resonant point, and so
a small $\lambda$ value can yield the correct relic density.
From Eq.~(\ref{csel}), it is evident that the darkon-nucleon cross-section becomes small if
$\lambda$ is small.
It follows that direct detection of the darkon is not possible in the vicinity of
the resonant point.
In contrast, as one can see from Fig.~\ref{cross-sec-md}, away from resonant point 
the cross-section can be large enough to be measurable.
However, for given cross-section values, there may be two solutions for $m_D^{}$ with
$m_h^{}$ fixed, and sometimes there can be more than two solutions.
For example, taking \,$m_h^{}=120$\,GeV\, and the cross-section values from the new
upper-limit from CDMS\,II, we find that the darkon mass can be about \,53\,GeV or~74\,GeV.\,
There can be more ambiguities in other cases.
For example, with \,$m_h^{}=120$\,GeV\, again, if a cross-section of order
\,$2\times 10^{-44}$\,cm$^2$\, is measured, the value of $m_D^{}$ can be \,55, 67,
or~79~GeV.\,
Hence, even if the LHC can obtain the Higgs mass, the darkon mass may be determined
only up to some discrete ambiguities.

Without a direct measurement of the darkon mass, one may be able to resolve some of these
discrete ambiguities if the branching ratio of the Higgs decaying into invisible channels
can be measured.  This would certainly work in the case of a twofold ambiguity.
Since the darkon is stable, the darkon pairs produced in the decay mode  \,$h\to DD$\, will
be invisible.  If $m_h^{}$ is larger than $2m_D^{}$, this new channel becomes open,
which increases the Higgs invisible branching-ratio.
On the other hand, if $m_h^{}$ is smaller than $2m_D^{}$, the Higgs invisible
branching-ratio is just that in the SM and not affected by the introduction of the darkon.

An alternative situation arises if the darkon mass is found first from a direct-search
experiment.  In this case, after measuring the darkon-nucleon cross-section, one may also
encounter some discrete ambiguities in establishing the Higgs mass, as illustrated
in Fig.~\ref{cross-sec-mh}.
For instance, assuming that \,$m_D^{}=70$\,GeV\, and the cross-section is
\,$3.8\times 10^{-44}$\,cm$^2$,\, just like the CDMS\,II upper-limit, one finds from this
figure the Higgs mass to be \,$m_h^{}\simeq119$\,GeV\, or \,178\,GeV.\,
To resolve the ambiguity, one would have to determine the Higgs mass at a collider.
From all these considerations, it is clear that LHC measurements of the Higgs mass and
invisible decay modes will yield crucial complementary information about the SM+D.

\begin{figure}[t]
\includegraphics[width=3.2in]{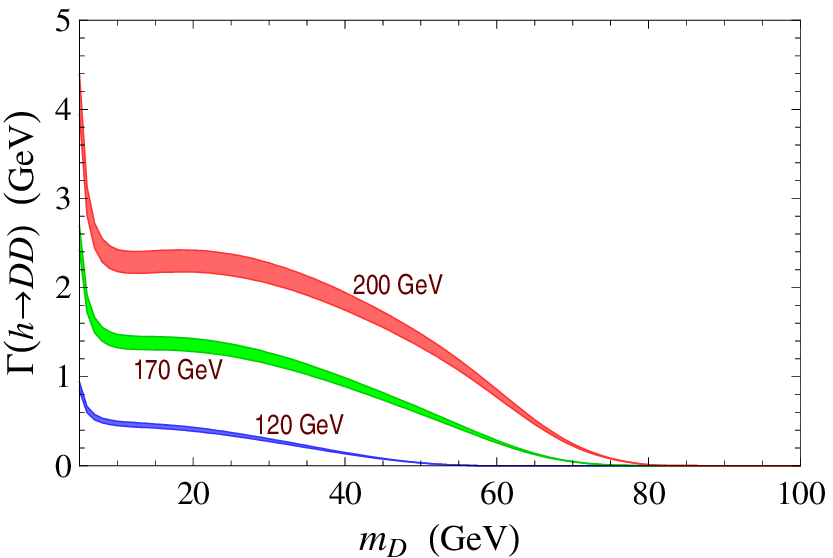} \, \,
\includegraphics[width=3.2in]{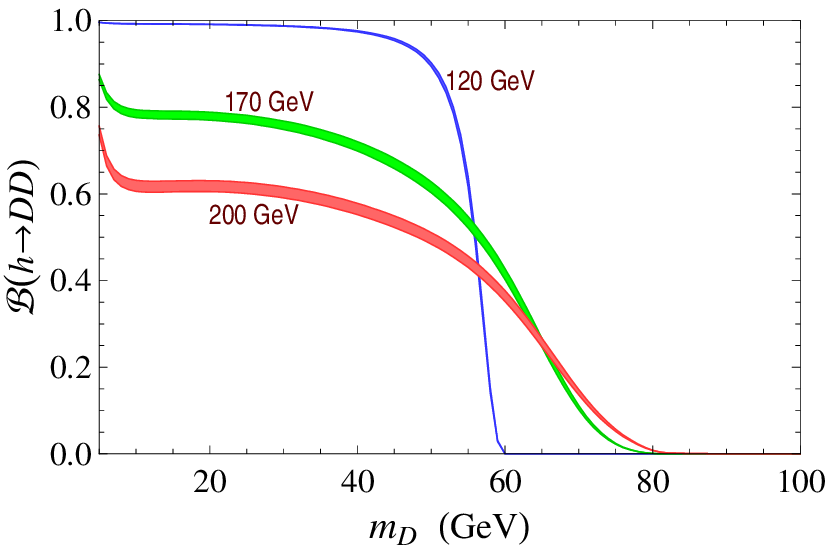}
\caption{Partial width and branching ratio of the invisible decay \,$h\to DD$\, as functions
of the darkon mass $m_D^{}$ for Higgs mass values  \,$m_h^{}=120,170,200$\,GeV.\,\label{br}}
\end{figure}

Finally, we turn to the effect of the darkon on the branching ratio
of the Higgs invisible decay, in relation to probing the darkon properties.
In fact, since the darkon interacts primarily with the Higgs
boson, its greatest impact is on the Higgs sector.
The existence of the darkon can give rise to enhancement of the Higgs width
via the additional process \,$h\to DD$\, if \,$m_h^{}>2m_D^{}$.\,
We have calculated the branching ratio of this invisible mode.
The results are depicted in Fig.~\ref{br}, where the mass choices are the same as those in
Fig.~\ref{relic-md}.
We observe that \,$h\to DD$\, can have a significant branching ratio.
This enhancement turns out to be advantageous.
For example, suppose that $m_h^{}$ is determined at the LHC to be 120\,GeV\,
and a DM measurement consistent with the CDMS\,II upper-limit is obtained.
Then, as illustrated earlier, there are two solutions for the darkon mass,
\,$m_D^{}=53$\,GeV and 74\,GeV.\,
The invisible decay mode at \,$m_D^{}=53$\,GeV\, is seen to dominate the Higgs decays,
with a branching ratio of order~80\%, whereas the invisible decay mode at \,$m_D^{}=74$\,GeV\,
is just that in the SM.
For a Higgs boson with a~large invisible branching fraction ($>$\,60\%) and a~mass within
the range
\,$120{\rm\,GeV}\mbox{\footnotesize\,$\lesssim$\,}m_h^{}\mbox{\footnotesize\,$\lesssim$\,}300$\,GeV,\,
direct Higgs searches at CMS through the usual SM modes may be unfeasible with 30\,fb$^{-1}$ of
integrated luminosity~\cite{Barger:2007im}.  However, with the same luminosity such a~Higgs
boson can be observed at ATLAS~\cite{Burgess:2000yq,He:2007tt,Barger:2007im,Davoudiasl:2004aj}.
Once the Higgs mass and invisible width are known from the LHC with sufficient precision,
one also knows to which mass region the darkon belongs, whether \,$2m_D^{}<m_h^{}$\, or
\,$2m_D^{}>m_h^{}$.\,
Direct DM search experiments can then focus on that particular range of $m_D^{}$ for
verification and providing further information on the darkon.
If \,$2m_D^{}<m_h^{}$\, and the Higgs invisible width is enlarged, it may also
be possible to infer the darkon mass from such plots as in Fig.~\ref{br}
and, in turn, the darkon-Higgs coupling $\lambda$ as well from a graph like Fig.~\ref{relic-md}.

In conclusion, we have studied some implications of the new results from the CDMS\,II
experiment for the simplest WIMP DM model, the standard model plus darkon.
We have found that the SM+D can offer a consistent interpretation of the CDMS\,II results,
with much of its parameter space still allowed by the data.
Since the model has a small number of parameters, there are strong correlations among
the darkon mass, darkon-nucleon cross-section, mass of the Higgs boson, and branching ratio
of its invisible decay.
Therefore, the interplay between direct searches for dark matter and the LHC study of
the Higgs boson can yield crucial information about the darkon properties.

\acknowledgments \vspace*{-3ex}
This work was partially supported by NSC, NCTS, and NNSF.

\end{document}